\documentclass[twocolumn,preprintnumbers,amsmath,aps]{revtex4}
\usepackage{graphicx}
\usepackage{dcolumn}
\usepackage{bm}
\usepackage{color}
\begin{document}
\def\eq#1{(\ref{#1})}
\def\fig#1{Fig.\hspace{1mm}\ref{#1}}
\def\tab#1{\hspace{1mm}\ref{#1}}
\title{Anisotropy effects on the critical magnetic field in CaC$_{6}$ superconductor}  
\author{Ewa A. Drzazga$^{\left(1\right)}$}
\email{edrzazga@wip.pcz.pl}
\author{Iza A. Wrona$^{\left(2\right)}$}
\author{Rados{\l}aw Szcz{\c{e}}{\'s}niak$^{\left(1\right)}$}
\affiliation{$^1$ Institute of Physics, Cz{\c{e}}stochowa University of Technology, Ave. Armii Krajowej 19, 42-200 Cz{\c{e}}stochowa, Poland}
\affiliation{$^2$ Institute of Physics, Jan D{\l}ugosz University in Cz{\c{e}}stochowa, Ave. Armii Krajowej 13/15, 42-200 Cz{\c{e}}stochowa, Poland}
\date{\today}
\begin{abstract}

The calcium intercalated graphite (CaC$_{6}$) is considered to be a representative material of the graphite intercalated superconductors, which exhibits sizable anisotropy of the Fermi surface (FS). Herein, the influence of the FS anisotropy on the critical magnetic field ($H_{C}$) in CaC$_{6}$ superconductor is comprehensively analyzed within the Migdal-Eliashberg (M-E) formalism. To precisely account for the mentioned anisotropy effects, the analysis is conducted in the framework of the six-band approximation, hitherto not employed for calculations of the $H_{C}$ function in CaC$_{6}$ material. For convenience, the obtained results are compared with the available one- and three-band estimates reported by using the M-E theory. A notable signatures of the increased number of bands are observed for the temperature dependent $H_{C}$ functions. In particular, the $H_{C}$ function decreases at T=0 K as the number of the considered bands is higher. Moreover, it is argued that the six-band formalism yields the most physically relevant shape of the $H_{C}$ function among all considered approximations. Therefore, the six-band model introduces not only quantitative but also qualitative changes to the results, in comparison to the other discussed FS approximations. This observation supports postulate that the six-band model constitutes minimal structure for FS of CaC$_{6}$, but also magnify importance of anisotropy effects for the critical magnetic field calculations.

\end{abstract}
\maketitle
{\bf Keywords:} ${\rm CaC_{6}}$ compound, multi-band superconducting state, thermodynamic properties, strong-coupling formalism. 
\vspace*{0.5cm}
%

\section{INTRODUCTION}

Graphite, the most thermodynamically stable of carbon allotrope forms, is a system of layered structure with a notable difference between the in- and out-of-plane interactions. Specifically, the interactions between the atoms in the same plane are stronger than interactions between atoms in the neighboring planes; the carbon-carbon atoms are connected by the strong covalent bonds ($\pi_{z}$), whereas graphene layers are brought togheter by the weak Van der Waal's bonds ($\sigma$). Such presence of two types of bonds results in a highly anisotropic properties of graphite \cite{Emery2008A}. Because the carbon $\sigma$ bands are completely filled, the $\pi$ bands have dominant character on the Fermi surface of carbon \cite{Sanna2007A}. At the same time, from this quasi-bidimensionality arises a large number of other interesting properties, namely: the differences between elastic constants parallel and perpendicular to the graphite planes, the different conductivities {\it etc.} \cite{Lang1994A}.

Unfortunately, the pure graphite does not exhibits superconducting properties \cite{Ginzburg1964A,Kudasov2006A}, and the introduction of the foreign metal atoms between graphite layers (known as the intercalation process) is required to induce superconducting condensate \cite{Lang1994A,Calandra2005A,Kim2007A}. After intercalation the structure of graphite within a given carbon layer does not change much \cite{Weller2005A}, however the inter-planar distances and surface plane aggregation are the subject to modifications \cite{Emery2005A}. Nowadays, over 100 reagents are know which can intercalate graphite \cite{Dresselhaus2002A}, with calcium being the most prominent example. In particular, the CaC$_{6}$ compound exhibits the highest value $T_{C}=11.5$~K among all graphene intercalated compounds (GICs) \cite{Weller2005A}. Yet another characteristic feature of superconducting state in CaC$_{6}$ is the fact that its Fermi surface is highly anisotropic \cite{Massidda2009A} and therefore requires multi-band theoretical models for the appropriate description \cite{Massidda2009A, Szczesniak2015A}. 

The aim of presented work is to discuss the influence of Fermi surface (FS) anisotropy on the critical magnetic field ($H_{C}$) in CaC$_{6}$ superconductor. The motivation for these investigations are the previous findings which suggest appearance of a notable anisotropy signatures in the properties of CaC$_{6}$, such as: the pairing gap, the electron-phonon and electron-electron interactions, as well as the electron effective mass \cite{Sanna2007A, Massidda2009A, Szczesniak2015A}. In this manner our work appears as a supplementary to the already available studies, but also attempts to visualize discussed effects in the magnetic features of CaC$_{6}$. Such complementary picture may be additionally of great importance for future studies on GICs as well as their purely two-dimensional counterparts like the lithium-decorated graphene \cite{Profeta2012A, Szczesniak2014J, Szczesniak2015D, Szczesniak2018C}, which is also predicted to present sizable anisotropy \cite{Ludbrook2015A, Zheng2016A}. To this end, present work attempts to verify the postulate that six-band model of FS in CaC$_{6}$ should be considered as a minimal approximation \cite{Sanna2007A}.

Specifically, the analysis of the aforementioned properties is conducted within the six-band Migdal-Eliashberg (M-E) formalism \cite{Migdal1958A, Eliashberg1960A}, hitherto not employed for the calculations of the $H_{C}$ function in CaC$_{6}$. The choice of the theoretical model is reinforced not only by the strong FS anisotropy but also by the relatively high electron-phonon coupling constant $\lambda=0.831$ in CaC$_{6}$, which prevents its description within the canonical Bardeen-Cooper-Schrieffer theory \cite{Bardeen1957A, Bardeen1957B, Cyrot1992A}. Finally, for convenience, the obtained results are compared with the predictions of the one- and three-band M-E model, reported previously in \cite{Szczesniak2015A}. 


\section{Theoretical model}

\begin{figure}
\includegraphics[width=\columnwidth]{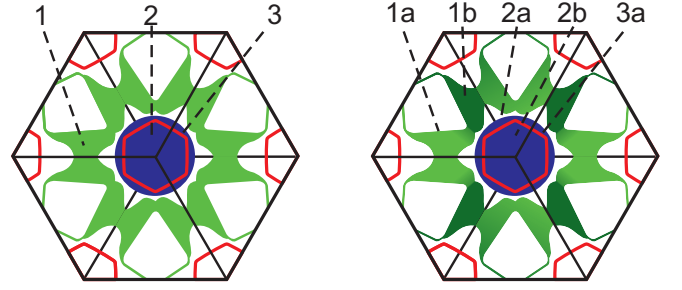}
\caption{Graphical representation of the band indexes within the three- and six-band approximation of Fermi surface for the CaC$_{6}$ compound \cite{Massidda2009A}.}
\label{f1}
\end{figure}

As mentioned above, the thermodynamic properties of interest for the superconducting state in ${\rm CaC_{6}}$ are analyzed here within the multi-band Migdal-Eliashberg equations. The solutions of such equations are required for further estimation of the $H_{C}$ function in CaC$_{6}$. Specifically, due to the assumed six effective bands \cite{Sanna2012A}, the Eliashberg equations for the order parameter $(\Delta^{\alpha}_{n}=\Delta^{\alpha}(i\omega_{n}))$ and the wave function renormalization factor $(Z^{\alpha}_{n}=Z^{\alpha}(i\omega_{n}))$ on the imaginary axis take the following form:
\begin{eqnarray}
\label{r1}
\nonumber
&&\Delta^\alpha_{n}Z^\alpha_{n} = \pi k_{B}T\\
&& \sum_{\beta}\sum^{M}_{m=-M}
\frac{[K^{\alpha\beta}\left(\omega_{n}-\omega_{m}\right)
-\mu^{\star}_{\alpha\beta}\left(\omega_{m}\right)]}
{\sqrt{\omega^{2}_{m}+\left(\Delta^\beta_{m}\right)^2}} 
{\Delta^\beta_{m}},
\end{eqnarray}
and
\begin{equation}
\label{r2}
Z^\alpha_{n}=1+\pi k_{B}T\sum_{\beta}\sum^{M}_{m=-M}\frac{K^{\alpha\beta}
\left(\omega_{n}-\omega_{m}\right)}{\sqrt{\omega^{2}_{m}+
\left(\Delta^\beta_{m}\right)^{2}}}\frac{\omega_{m}}{\omega_{n}} Z^{\beta}_{m},
\end{equation}
where symbols $\alpha,\beta\in\left\{1a, 1b, 2a, 2b, 3a, 3b\right\}$ denote the index of the electronic band. The assumed division arises from the previously defined characteristic Fermi surface points within the three-band approximation (see Fig. \ref{f1} (A)) \cite{Massidda2009A}. In particular, the physically distinct regions are: the outer $\pi$ FS part (band 1), the spherical FS part (band 2) and the $\pi$ FS part crossing the spheres (band 3). It turns out that each of the mentioned parts can be further divided into the two coupled parts named {\it a} and {\it b}, resulting in the employed six-band approximation \cite{Sanna2012A}.

Moreover, in Eq. (\ref{r1}) and (\ref{r2}) the fermion Matsubara frequency is defined by: $\omega_{n}=\pi k_{B}T\left(2n-1\right)$, where $k_{B}$ represents the Boltzmann constant. In order to obtain the numerical stability of our calculations we assume 300 Matsubara frequencies. Such assumption allows us to obtain physically relevant and stable results for $T>T_{0}$ where $T_{0}=2$ K. Finally, the $K^{\alpha\beta}(z)$ function is the electron-phonon pairing kernel:
\begin{equation}
\label{r3}
K^{\alpha\beta}(\omega_{n}-\omega_{m})=\lambda^{\alpha\beta}\frac{\Omega^{2}_{C}}
{(\omega_{n}-\omega_{m})^{2}+\Omega^{2}_{C}},
\end{equation}
where $\lambda^{\alpha\beta}$ denotes the electron-phonon coupling constant, which is defined in the matrix form as follows \cite{Sanna2012A}:
\begin{equation}
\label{r4}
\left[\lambda^{\alpha\beta}\right]=\left[\begin{array}{cccccc} 
0.163 & 0.126 & 0.099 & 0.033 & 0.201 & 0.046 \\
0.179 & 0.140 & 0.105 & 0.035 & 0.221 & 0.050 \\
0.331 & 0.245 & 0.151 & 0.084 & 0.384 & 0.096 \\
0.271 & 0.202 & 0.206 & 0.047 & 0.400 & 0.080 \\
0.252 & 0.194 & 0.145 & 0.061 & 0.309 & 0.073 \\
0.206 & 0.157 & 0.128 & 0.044 & 0.259 & 0.060
\end{array}\right]. 
\end{equation}
The Eliashberg equations (\ref{r1}) and (\ref{r2}) are numerically solved by using the modified numerical procedures initially developed for the isotropic materials \cite{Szczesniak2017A, Szczesniak2015F, Szczesniak2014I, Szczesniak2018A, Szczesniak2018B} and later adopted to three-band CaC$_{6}$ superconductor \cite{Szczesniak2015A}. In this context, the input parameters for the calculations are the critical temperature $T_{C}=11.5$ K and the corresponding elements of the Coulomb pseudopotential ($\mu^{\star}_{\alpha\beta}$) matrix \cite{Sanna2012A}: 
\begin{equation}
\label{r5}
\left[\mu^{\star}_{\alpha\beta}\right]=
\left[\begin{array}{cccccc} 
 0.250 & 0.250 & 0.250 & 0.250 & 0.250 & 0.250 \\
 0.176 & 0.176 & 0.176 & 0.176 & 0.176 & 0.176 \\
 0.075 & 0.075 & 0.075 & 0.075 & 0.075 & 0.075 \\
 0.031 & 0.031 & 0.031 & 0.031 & 0.031 & 0.031 \\
 0.200 & 0.200 & 0.200 & 0.200 & 0.200 & 0.200 \\
 0.056 & 0.056 & 0.056 & 0.056 & 0.056 & 0.056
\end{array}\right].
\end{equation}
The columns and rows of Eq. (\ref{r4}) and (\ref{r5}) are arranged with respect to the assumed band indices. It is also worth to note, that the $\mu^{\star}$ parameter in the one-band case is equal to $0.21$ \cite{Sanna2012A, Massidda2009A}, and this value corresponds to the only non-zero eigenvalue of the matrix $\left[\mu^{\star}_{\alpha\beta}\right]$. 

\begin{figure}
\includegraphics[width=\columnwidth]{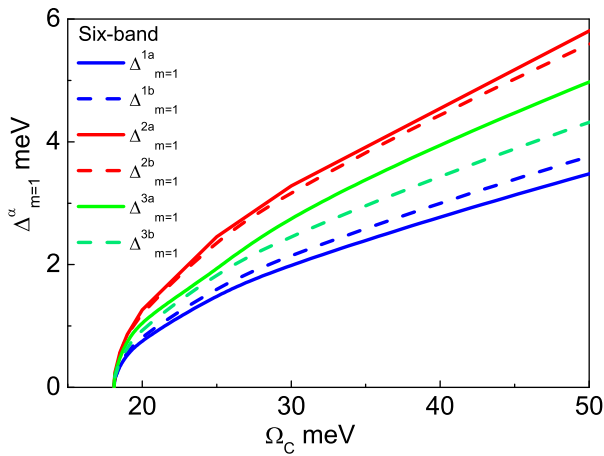}
\caption{The maximum value of the order parameter as a function of the characteristic phonon frequency at $T=T_{C}$ for CaC$_{6}$, as analyzed within the six-band Eliashberg equations.}
\label{f2}
\end{figure}

Next, by using the explicit forms of matrices $\left[\lambda^{\alpha\beta}\right]$ and $\left[\mu^{\star}_{\alpha\beta}\right]$, the physical value of the characteristic phonon frequency $\Omega_{C}$ has to be estimated. For this purpose the set of equations of the following form is solved: $\left[\Delta^{\alpha}_{n=1}\left(\Omega_{C}\right)\right]_{T=T_{C}}=0$. The obtained results are depicted in Fig. \ref{f2}. Because all the bands are connected with each other, the $\Delta^{\alpha}_{n=1}$ value tends to be zero at the value of $\Omega_{C}=18.11$~meV. For comparison, in the three-band case this value is equal to $\Omega_{C}=20.21$~meV \cite{Szczesniak2015A}. The results presented in Fig. \ref{f2} are used for the, pivotal to this work, calculations of the $H_{C}$ functions in CaC$_{6}$, as presented in the next section.

\section{Numerical results}

\begin{figure}
\includegraphics[width=\columnwidth]{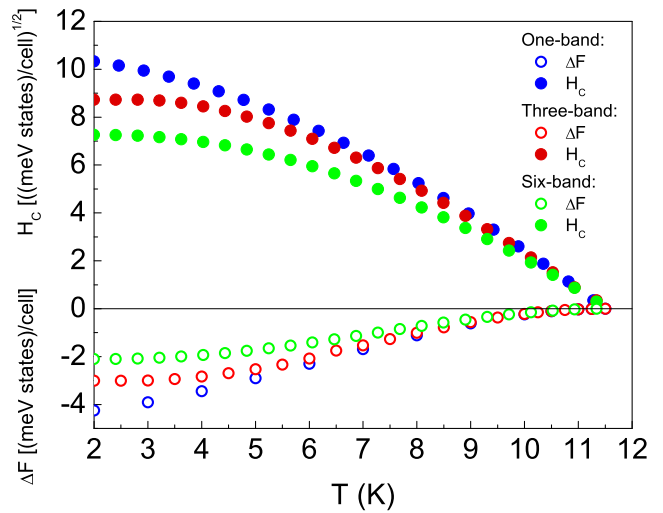}
\caption{The critical magnetic field (upper panel) and the free energy difference between the superconducting and normal state (lower panel) for CaC$_{6}$, as obtained within the six-band formalism. For comparison results for the one- and three-band approximation, adopted from \cite{Szczesniak2015A}, are depicted.}
\label{f3}
\end{figure}

The critical magnetic field ($H_{C}$) is defined in the following form, suitable for the Eliashberg formalism analysis \cite{Carbotte1990A}:
\begin{eqnarray}
\label{r6}
H_{C}=\sqrt{-8\pi\Delta F}.
\end{eqnarray}
where $\Delta F$ denotes the free energy difference between the normal ($N$) and superconducting state ($S$) in the considered material. The $\Delta F$ function relates the critical magnetic field to the solutions of the Eliashberg equations from the previous section. In particular, the $\Delta F$ function can be written for the six-band approximation in the following way \cite{Bardeen1964A}: 
\begin{eqnarray}
\label{r7}
\Delta F&=&-2\pi k_{B}T\sum^{M}_{m=1}\sum_{\alpha\in\{a,b,c\}}\rho_\alpha\left(0\right)\\ \nonumber
&\times&
[\sqrt{\omega^2_m+\left(\Delta^\alpha_m\right)^2}-|\omega_m|]\\ \nonumber
&\times&
[Z^{\alpha,\left(S\right)}_m-Z^{\alpha,\left(N\right)}_m \frac{|\omega_m|}{\sqrt{\omega^2_m+\left(\Delta^\alpha_m\right)^2}}],
\end{eqnarray}
were the $Z^{\alpha,\left(S\right)}$ and $Z^{\alpha,\left(N\right)}$ denotes the wave function renormalization factor for the superconducting ($S$) and normal ($N$) state, respectively. Furthermore, the symbol $\rho_\alpha\left(0\right)$ constitute the vector of the electron density of states in the six-band formalism \cite{Sanna2012A}: 
\begin{equation}
\label{r8}
\left[\rho_{\alpha}\left(0\right)\right]=\left[\begin{array}{c}       0.250 \\
                                                                      0.031 \\
                                                                      0.176 \\
                                                                      0.200 \\
                                                                      0.075 \\
                                                                      0.056
\end{array}\right] \left[{\rm \frac{states}{eV*cell}}\right], 
\end{equation}
with $\rho\left(0\right)=\sum_{\alpha}\rho_{\alpha}\left(0\right)$. It is important to note that the $\Delta F$ function is not only a input parameter for the critical magnetic field calculations, but also allows to verify the physical relevancy of the obtained results. For this reason we plot the $\Delta F$ as a function of temperature in the lower panel of Fig. \ref{f3}. The results determined within the six-band approximation are plotted by using the open green circles. Additionally, the one- and three-band estimates, obtained in \cite{Szczesniak2015A}, are given in the lower panel of Fig. \ref{f3} and marked by the blue and red open circles, respectively. One can observe notable differences between three presented cases, specifically, when the number of bands increases the value of the $\Delta F$ function decreases in the range of low temperatures. Obviously, together with the increase of the temperature all considered cases converge, due to the fact that each of the sets has been obtained for the same value of $T_{C}=11.5$ K. Nonetheless, observed differences suggest the saturation of the numerical calculations together with the increasing number of the bands included in the employed approximation. Finally, from the physical point of view, the six-band results obtained herein confirm the thermodynamic stability of superconducting state for $T \in (T_{0}, T_{C})$, by exhibiting negative values in accordance to the one- and three-band cases.

In the same figure, {\it i.e.} Fig. \ref{f3}, the calculated $H_{C}$ as a function of temperature is depicted in the upper panel. In contrary to the $\Delta F$ function our estimates for the $H_{C}$ are marked by the closed green points. Once again the results determined previously in \cite{Szczesniak2015A} are marked for comparison by the closed blue and red circles for the one- and three-band model, respectively. Similarly to the predictions given in the lower panel a sizable effect of the anisotropy on the obtained $H_{C}$ results is noted. In particular, when the number of the bands included in the approximation increases the $H_{C}$ function for the low temperatures decreases. However, the most pivotal observation can be made by closer inspection of the $H_{C}$ function shape for different approximations. One can see that the $H_{C}$ parameter obtained within the six-band formalism gives almost linear decrease with the increasing temperature above $T \sim 7$ K. This is in agreement with the results adopted from \cite{Szczesniak2015A}. Nonetheless, only results derived from the six-band Eliashberg equations present characteristic plateau for the lower temperatures, which is the most physically relevant behavior of $H_{C}$ for such range of the temperature.
   
\section{Summary}

In the present paper the analysis of the critical magnetic field in CaC$_{6}$ was conducted within the six-band Migdal-Eliashberg equations. For convenience, the obtained results were compared to the previous estimates made in the framework of the one- and three-band M-E formalism. In particular, the conducted computations predict that the six-band approximation gives much lower values of $H_{C}$ in the range of low temperatures, suggesting saturation of the Eliashberg theory solutions together with the increase of number of bands included in the employed model. Moreover, within the same temperature range, the $H_{C}$ function exhibits a characteristic plateau-like behavior. Such behavior is argued to be most physically relevant among all discussed approximations. In what follows, it is suggested that the anisotropy in CaC$_{6}$ strongly influences the $H_{C}$ function, not only qualitatively but also quantitatively. Finally, determined estimates also support postulate that the six-band approximation is the minimal model for the proper description of the FS in CaC$_{6}$.

\begin{acknowledgments}
E. A. Drzazga would like to acknowledge financial support of this work under Cz{\c e}stochowa University of Technology Research Grant for Young Scientists (grant no. BS/MN-203-301/2018).
\end{acknowledgments}

\bibliography{Bibliography}
\end{document}